\documentclass[conference]{IEEEtran}
\usepackage{graphicx}
\usepackage{float} 
\usepackage{url} 
\usepackage{color}

  \usepackage{caption}
  \usepackage{subcaption}
  \usepackage{setspace}

\ifCLASSINFOpdf
\else
\fi
\begin{document}
%
\title{Sensor Discovery and Configuration Framework for The Internet of Things Paradigm}

\author{\IEEEauthorblockN{Charith Perera, Prem Prakash Jayaraman,\\ Arkady Zaslavsky,  Dimitrios Georgakopoulos\\ }
\IEEEauthorblockA{CSIRO Computational Informatics, Canberra, ACT 2601, Australia\\
\{charith.perera, prem.jayaraman, arkady.zaslavsky, \\  dimitrios.georgakopoulos\}@csiro.au}
\and
\IEEEauthorblockN{Peter Christen}
\IEEEauthorblockA{Research School of Computer Science,\\ The Australian National University, \\Canberra, ACT 0200, Australia\\
peter.christen@anu.edu.au}
}


%


\maketitle

\begin{abstract}
Internet of Things (IoT) will comprise billions of devices that can sense, communicate, compute and potentially actuate. The data generated by the Internet of Things are valuable and have the potential to drive innovative and novel applications. The data streams coming from these devices will challenge the traditional approaches to data management and contribute to the emerging paradigm of big data. One of the most challenging tasks before collecting and processing data from these devices (e.g. sensors) is discovering and configuring the sensors and the associated data streams. In this paper, we propose a tool called \textit{SmartLink} that can be used to discover and configure sensors. Specifically, \textit{SmartLink}, is capable of discovering sensors deployed in a particular location despite their heterogeneity (e.g. different communication protocols, communication sequences, capabilities). \textit{SmartLink} establishes the direct communication between the sensor hardware and cloud-based IoT middleware. We address the challenge of heterogeneity using a plugin architecture. Our prototype tool is developed on the Android platform. We evaluate the significance of our approach by discovering and configuring 52 different types of Libelium sensors.

\end{abstract}


%
\IEEEpeerreviewmaketitle

\section{Introduction}
\label{sec:Introduction}

The Internet of Things (IoT) received its first attention in late 20th century. The term was coined in 1998 \cite{P065} and later defined as \textit{``The Internet of Things allows people and things\footnote{We use  terms, `\textit{objects}' and `\textit{things}' interchangeably to give the same meaning as they are frequently used in the IoT related documentation. Other terms used by the research community are `smart objects', `devices', `nodes'.} to be connected Anytime, Anyplace, with Anything and Anyone, ideally using Any path/ network and Any service''} \cite{P019}. As highlighted in the  definition, connectivity among devices is a critical functionality that is required to fulfil the vision of the IoT. The following statistics highlight the magnitude of the challenge we need to address in the future and motivate our research. Due to increasing popularity of mobile phones over the past decade, it is estimated that there about 1.5 billion Internet-enabled PCs and over 1 billion Internet-enabled  mobile phones  today. The number of things connected to the Internet has exceeded the number of people on earth in 2008. By 2020, there will  be 50 to 100 billion devices connected to the Internet \cite{P029}. Similarly, according to BCC Research \cite{P255}, the global market for sensors was around \$56.3 billion in 2010. In 2011, it has risen to around \$62.8 billion. Global market for sensors is expected to increase to \$91.5 billion by 2016, at a compound annual growth rate (CAGR) of 7.8\%.

Sensing as a service \cite{ZMP008} has also gained popularity among academia and industry. It envisions to offer sensing capability as a service similar to other offerings such as infrastructure as a service (IaaS), platform as a service (PaaS), and software as a service (SaaS). In such environment, discovering, connecting and configuring sensors\footnote{Each device may comprise one or more sensors. Such device can also be called as \textit{sensor node}. In this work, we configure the entire sensor node.} is critical so  cloud-based IoT platforms can retrieve data from sensors. The sensing as  a service paradigm and the importance of sensor configuration is further discussed in \cite{ZMP008}. This work is also motivated by our previous work which focused on utilising mobile phones and similar capacity devices to collect sensor data. In DAM4GSN \cite{ZMP001}, we proposed an application that can be used to collect data from sensors built-in to  mobile phones. Later, we proposed MoSHub \cite{ZMP005} that allows a variety of different external sensors to be connected to a mobile phone using an extensible plugin architecture. MoSHub also configures the cloud middleware accordingly. In MOSDEN \cite{ZMP009}, we developed a complete middleware for resource constrained mobile devices that is capable of collecting data from both internal and external sensors. MOSDEN can also apply SQL-based fusing on data streams in real-time. As we mentioned earlier, \textit{in order to collect data from sensors, first we need to discover and configure the available sensors in such a way a that cloud can communicate with them}. In our previous work, discovery and configuration steps are performed manually. In this work, we propose an approach that can be used to discover and configure sensors autonomously.

\section{Research Challenges}
\label{sec:Research Challenges}

A \textit{sensor configuration process} detects, identifies, and configures sensor hardware  and cloud-based IoT platforms in such a way that software platforms can retrieve data from sensors when required. In this section, we identify the importance of sensor configuration, several major challenges and factors that need to be considered. The process of sensor configuration in the IoT is important due to two main reasons.  Firstly, it establishes the connectivity between sensor hardware and software systems which allows to retrieve data from sensor. Secondly, it allows to optimize the sensing and data communication by considering several factors as discussed below. Let us consider the following problem \textit{`Why is sensor configuration a challenging task in the IoT environment?'}. The major factors that makes sensor configuration challenging are \textit{1) the number of sensors, 2) heterogeneity, 3) scheduling, sampling rate, communication frequency, 4) data acquisition, 5) dynamicity,} and \textit{6) context} \cite{ZMP007}.



\noindent\textbf{1) Number of Sensors:} When the number of sensors that need to be configured is limited, we can use manually or semi-autonomous techniques. However, as the numbers grow rapidly towards millions and billions, such methods become extremely inefficient, expensive, labour-intensive, and in most situations impossible. Therefore, large numbers have made sensor configuration challenging. An ideal sensor configuration approach should be able to configure sensors autonomously as well as within very short periods of time.

\begin{figure}[h]
 \centering
 \includegraphics[scale=0.30]{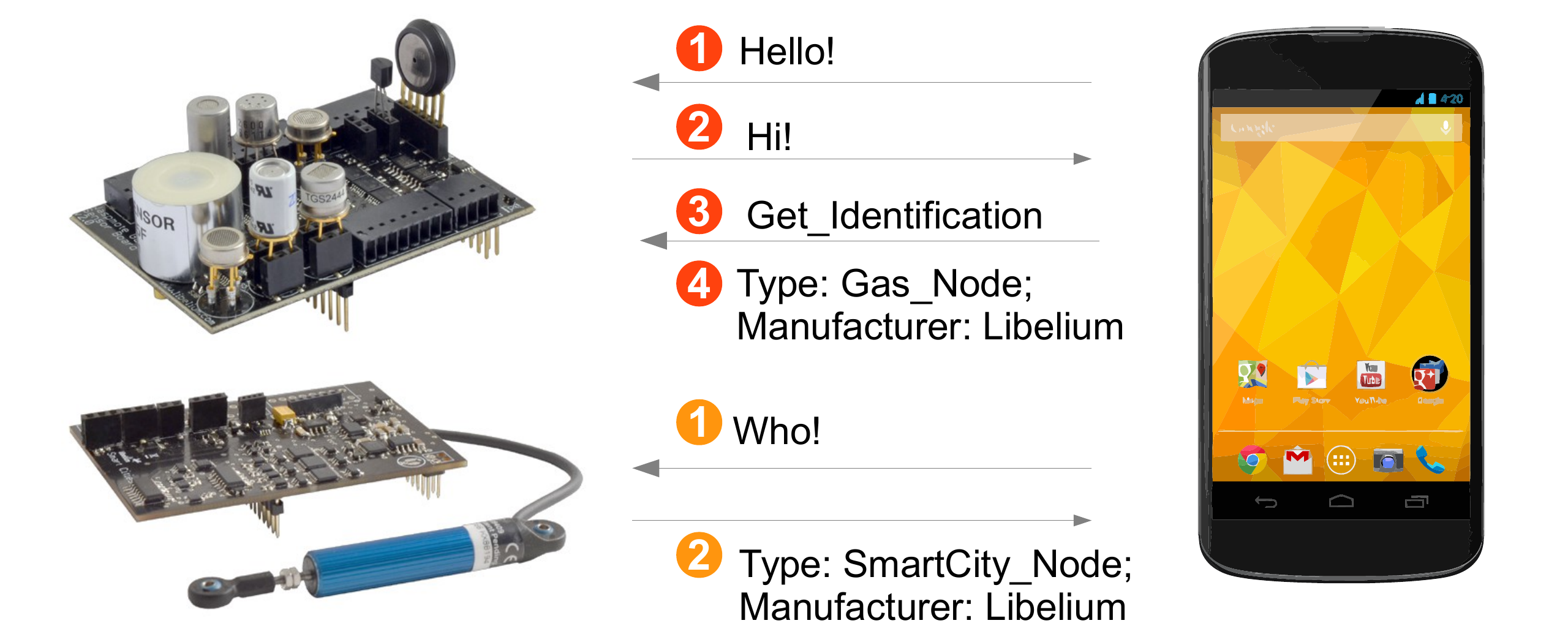}
 \caption{Heterogeneity in term of communication and message/command passing sequences. Some sensors may need only a few message/command passes and others may require more. The messages/commands understood by each sensors may also vary.}
 \label{Figure:Communication_Sequence}
\vspace{-8pt}
\end{figure}

\noindent\textbf{2) Heterogeneity:} This factor can be interpreted in different perspectives. Firstly, heterogeneity in term of communication technologies used by the sensors as presented in Table \ref{Table:Wireless Technology Comparison}. Secondly, the heterogeneity in term of measurement capabilities (e.g. temperature, humidity, motion, pressure). Thirdly, the types of data (e.g. numerical (small in size), audio, video (large in size)) generated by sensors are also heterogeneous. Finally, the communication sequences, as depicted in Figure \ref{Figure:Communication_Sequence}, and security mechanisms used by different sensors, are also heterogeneous (e.g. exact messages and sequence need to be followed to successfully communicate with a given sensor).  These differences make the sensor configuration process challenging. An ideal sensor configuration approach that is designed for the IoT paradigm should be able to handle such heterogeneity. Towards this, the proposed solution should be scalable and  should provide support for new sensors as they come to the market.

\begin{table}[htbp]
\footnotesize
\centering
\vspace{-8pt}
\caption{Wireless Technology Comparison}
\begin{tabular}{@{}p{2.42cm}@{} p{1.6cm}@{} p{1.1cm}p{1.4cm} @{}p{1.5cm}}
\hline 
 & \textbf{ZigBee} & \textbf{GPRS-GSM} & \textbf{WiFi} & \textbf{Bluetooth} \\ \hline \hline 
Standard & 802.15.4 &  & 802.11b & 802.15.1 \\ 
System Resources & 4-32KB & 16MB+ & 1MB+ & 250KM+ \\ 
Batterylife (days) & 100-1000+ & 1-7 & 0.5-5 & 1-7 \\ 
Network Size & $2^{64}$ & 1 & 32 & 7 \\ 
Bandwidth (KB/s) & 20-250 & 64-128+ & 11000 & 720 \\ 
Transmission \newline Range (meters) & 1-100+ & 1000 & 1-100 & 1-10+ \\ 
Success Metrics & Reliability, Power, Cost & Reach, Quality & Speed, Flexibility & Convenience, Cost \\ \hline
\end{tabular}
\label{Table:Wireless Technology Comparison}
\end{table}

\noindent\textbf{3) Scheduling, Sampling Rate, and Network Communication:} Sampling rate defines in which frequency sensors need to generate data (i.e. sense the phenomenon) (e.g. sense temperature every 10 seconds). Deciding an ideal (e.g. balance between user requirement and energy consumption) sampling rate can be a very complex task that has a strong relationship with \textbf{\textit{6) Context}}. Schedule defines the time table of sensing and data transmitting (e.g. sense temperature only between 8am to 5pm weekdays). Network communication defines the frequency of data transmission (e.g. send data to the cloud-based IoT platform every 60 seconds). Designing efficient sampling and scheduling strategies and configuring  sensors accordingly is challenging. 



\noindent\textbf{4) Data Acquisition:} This can be divided into two  categories: based on responsibility and frequency \cite{ZMP007}. There are two methods that can be used to acquire data from a sensor based on responsibility: push (e.g. cloud requests data from a sensor and the sensor responds with data) and pull (e.g. sensor pushes data to a cloud without explicit cloud request). Further, based on frequency, there are two data acquisition methods: instant (e.g. send data to the cloud when a predefined event occurs) and interval (e.g. periodically send data to the cloud). Pros, cons, and applicabilities are discussed in \cite{ZMP007}. Using appropriate data acquisition methods based on the context information is essential to ensure the efficiency.



\noindent \textbf{5) Dynamicity:} This means the frequency of changing  positions / appearing / disappearing of the sensors at a given location. IoT envisions that most of the objects we use in everyday lives will have sensors attached to them. Ideally, we need to connect and configure these sensors to software platforms in order to analyse data and understand the environment better. We observed several domains and broadly identified different levels of dynamicity based on mobility\footnote{It is important to note that same object can be classified into different levels depending on the environment they belongs to. Further, there is no clear definition to classify objects into different levels of dynamicity. However, our categorization allows to understand the differences in dynamicity.}. \textit{High level:} sensors move/ appear/ disappear at a higher frequency (e.g. RFID and other low-level, low-quality, less reliable, cheap sensors. Such sensors will be attached to consumables such as stationary, food packaging, etc.).  \textit{Low level:} sensors embedded and fitted into permanent structures (such as buildings and air condition systems) can be categorized under this level.  An ideal sensor configuration platform should be able to efficiently and continuously discover and re-configure sensors in order to cope up with high dynamicity.

\noindent\textbf{6) Context:} Context information plays a critical role in sensor configuration in the IoT. The objective of collecting sensor data is to understand the environment better by fusing and reasoning them. In order to accomplish this task, sensor data need to be collected in timely and location sensitive manner. Each sensor needs to be configured by considering context information. Let us consider a scenarios related to smart agriculture to understand why context matters in sensor configuration. \textit{Severe frosts and heat events can have a devastating affect on crops. Flowering time is critical for cereal crops and a frost event could damage the flowering mechanism of the plant. However, ideal sampling rate could be varied depending on both season of the year as well as the time of the day. For example, a higher sampling rate is necessary during the winter season and night time. In contrast, lower sampling would be sufficient during summer and day time. On the other hand, some reasoning approaches may required multiple sensor data reading. For example, a frost event can be detected by fusing air temperature, soil temperature, and humidity. However, if an air temperature sensor stopped sensing due to malfunctioning, there is no value in sensing humidity, because frost events cannot be detected without temperature. Therefore, configuring  the humidity sensor to sleep is ideal until the temperature sensor starts sensing again}. Such approaches save energy by eliminating unnecessary sensing and  communication.

\section{Architectural Design}
\label{sec:Architectural_Design}

\begin{figure}[h]
 \centering
\vspace{-1pt}
 \includegraphics[scale=1.1]{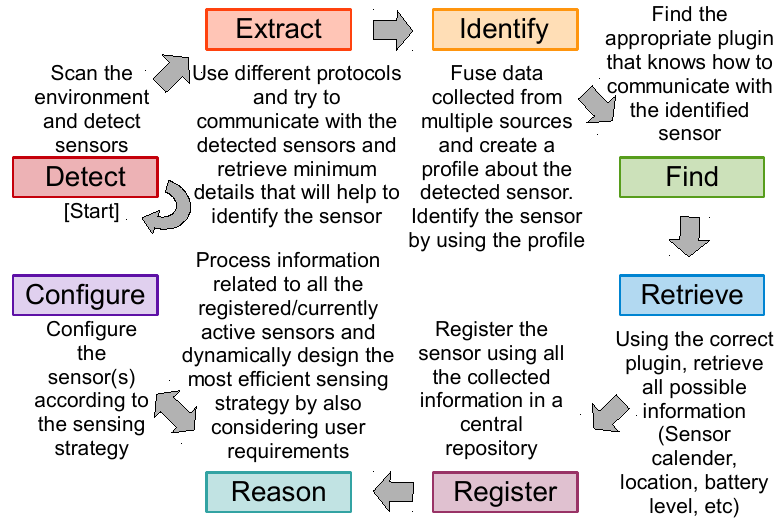}
 \vspace{-4pt}
 \caption{CADDOT Model for Sensor Configuration}
 \label{Figure:Model}
\vspace{-8pt}
\end{figure}

Previously, we identified several major factors that need to be considered when developing an ideal sensor configuration model for the IoT. This section presents the detailed explanation of our proposed solution: Context-aware Dynamic Discovery of Things (CADDOT). Figure \ref{Figure:Model} describes the main phases of the proposed model.

\noindent\textbf{Phases in CADDOT model:} The proposed model consists of eight (8) phases: \textit{detect, extract, identify, find, retrieve, register, reason,} and \textit{configure}. Some of the tasks mentioned in the model are performed by the \textit{SmartLink} tool and other tasks are performed by  cloud-based IoT middleware. 



\noindent\textit{\textbf{1) Detect:}} Sensors are configured to actively seek a wireless access points (WiFi or Bluetooth) which they can be connected to without any authorization, because at this point sensors do not have authentication details. Sensors will get them in \textbf{phase (8)}. \textit{SmartLink} becomes an open wireless hotspot so sensors can connected to it in an ad-hoc manner. 



\noindent \textit{\textbf{2) Extract:}} In this phase, \textit{SmartLink} extracts information from the sensor detected in the previous phase. Each sensor may be designed to respond to a different message-passing sequence depending on the sensor manufacturer and the sensor program developer. Even though the sensors and the \textit{SmartLink} may use the same communication technology/ protocol (e.g. TCP, UDP, Bluetooth), the exact communication sequence can be varied from one sensor to another. Therefore, it is hard to find the specific message passing sequence that each sensor follows. To address this challenge, we propose every sensor to respond to a common message during the communication initiation process. For example, \textit{SmartLink} broadcasts a message [WHO] where the sensors are expected to respond by providing a minimum amount of information about themselves, such as sensor unique identification number, model number / name, and manufacturer. This is similar to the TEDS mechanism discussed in \cite{P339}. It is important to note that we propose this [WHO] constraint only for minimum information extraction. Once the sensor is identified, subsequent communications and heterogeneity of  message-passing sequences is handled by matching plugins.

\noindent\textit{\textbf{3) Identify:}} \textit{SmartLink} sends all the information related to the newly detected sensor to the cloud. Cloud-based IoT middelware queries and reasons its data stores using the information. This process identifies the complete profile of the  sensor.

\noindent\textit{\textbf{4) Find:}} Once the cloud identifies the sensor uniquely, this information is used to find a matching plugin (can also be called driver) which knows how to communicate with a compatible sensor in full capacity. The IoT middleware pushes the plugin to \textit{SmartLink} where it gets installed\footnote{In practice, IoT middleware sends a request to \textit{Google Play store}. Google store pushes the plugin (i.e. an android application) to the \textit{SmartLink} autonomously via the Internet.}.

\noindent\textit{\textbf{5) Retrieve:}} Now, \textit{SmartLink} knows how to communicate with the detected sensor at full capacity with the help of the newly downloaded plugin. Next, \textit{SmartLink} retrieves the complete set of information that the sensor can provide (e.g. configuration details such as schedules, sampling rates, data structures /types generated by the sensor, etc.). Further, \textit{SmartLink} may communicate with other available sources (e.g. databases, web services) to retrieve additional information related to a sensor.

\noindent\textit{\textbf{6) Register:}} Once all the information about a given sensor has been collected, registration takes place in the cloud. The sensor descriptions are modelled according to the semantic sensor network ontology (SSNO) \cite{ZMP006}. This allows semantic querying and reasoning. Some of the performance evaluation related to the SSN ontology and semantic querying are presented in \cite{ZMP006}.

\noindent\textit{\textbf{7) Reason:}} This phase plays a significant role in the sensor configuration process.  It designs an efficient sensing strategy. Reasoning takes place in a distributed manner. The cloud IoT middleware retrieves data from sensors and other devices and identifies their availabilities and capabilities. Further, it considers the context information in order to design an optimized strategy. However, the technical details related to this reasoning process is out of the scope of this paper. At the end of this phase,  a comprehensive plan (e.g. schedule and sampling rate) for each individual sensor is designed.

\noindent\textit{\textbf{8) Configure:}} Sensors and cloud-based IoT software systems are configured based on the strategy designed in the previous phase. Schedules, communication frequency, and sampling rates that are custom designed for each sensor are  pushed into the  individual sensors. The connection between the sensor and the IoT software system are established  through direct wireless communication  or through intermediate devices such as MOSDEN \cite{ZMP009} so the cloud can retrieve data from sensors. The  details (IP address, port, authentication, etc.) required to accomplish above task is also provided to the sensor.

\section{Design Decisions and Applications}
\label{sec:Design_Decisions}

We made a number of design decisions during the development of the CADDOT model. These decisions address the challenges we highlighted in earlier sections.

\begin{figure}[t]
 \centering
 \includegraphics[scale=0.32]{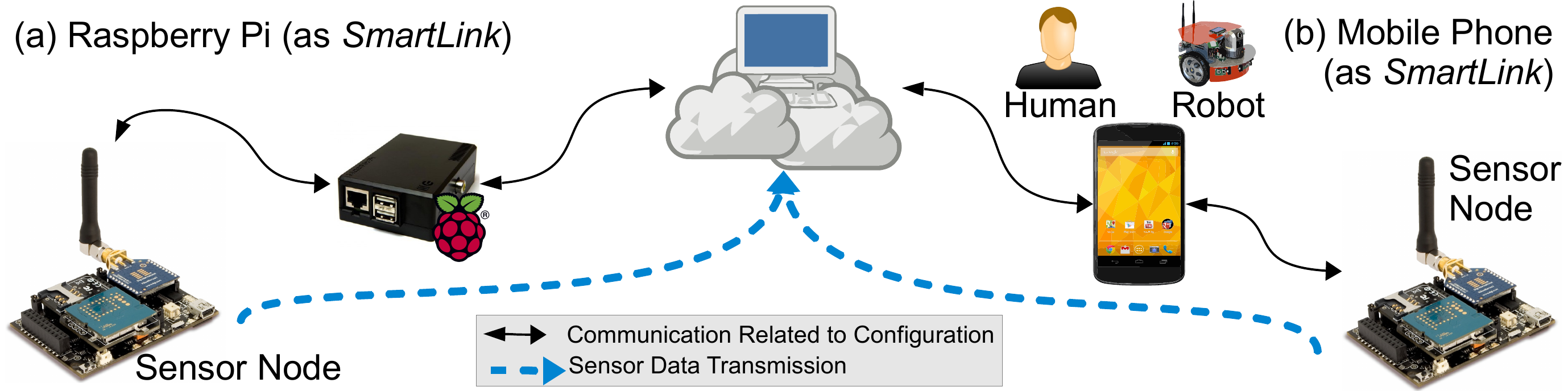}
 \caption{Application strategies of CADDOT model and \textit{SmartLink} tool. (a) usage of static \textit{SmartLink} (b) usage of mobile \textit{SmartLink}.}
 \label{Figure:Usage_Pattern}
\vspace{-18pt}
\end{figure}

\textbf{Security Concerns and Application Strategies:} There are different ways to employ our proposed model CADDOT as well as the tool \textit{SmartLink} in real world deployments. Figure \ref{Figure:Usage_Pattern} illustrates two different application strategies. It is important to note that neither our model nor the software tool is limited to a specific device or platform. In this paper, we conduct the experimentations on an Android-based mobile phone, as detailed in Section  \ref{sec:Implementation}. In   \textbf{\textit{ strategy (a)}}, a Raspberry Pi (raspberrypi.org) is acting as the \textit{SmartLink}  tool. This strategy is mostly suitable for smart home and office environments where WiFi is available. Raspberry Pi continuously performs the discovery and configuration process, as explained in Section  \ref{sec:Architectural_Design}. Finally, Raspberry Pi provides the authentication details to the sensor which is to connected to the secure home/office WiFi network. The sensor is expect to send data to the processing server (local or on cloud) directly over the secured WiFi network. In this strategy, \textit{SmartLink} is in static mode. Therefore, several \textit{SmartLink} instances installed on Raspberry Pi devices may be required to cover a building.  However, this strategy can handle a high level of dynamicity.

The \textbf\textit{{strategy (b)} }is  more suitable for situations where WiFi is not available or less dynamic. Smart agriculture can be considered as an example. In this scenario, sensors are deployed over a large geographical area (e.g. Phenonet: phenonet.com). Mobile robots (tractors or similar vehicles) with a \textit{SmartLink} tool attached to them can be used to discover and configure sensors. \textit{SmartLink} can then help to establish the communication between sensors and sinks. The permanent sinks used in the agricultural fields are usually low-level sinks (such as Meshlium \cite{P595}). Such sinks cannot perform sensor discovery or configuration in comparison to SmartLink. Such sinks are designed to collect data from sensors and upload to the cloud via 3G.


Other strategies can be built by incorporating the different characteristics pointed out in the above two strategies. This shows the extensibility of our solution. For example, Raspberry Pi, which we suggested for use as a \textit{SmartLink} in strategy (a), can be replaced by corporate mobile phones. So, without bothering the owner, the corporate mobile phones can silently perform the work of a \textit{SmartLink}.

\begin{figure}[t]
 \centering
 \includegraphics[scale=0.45]{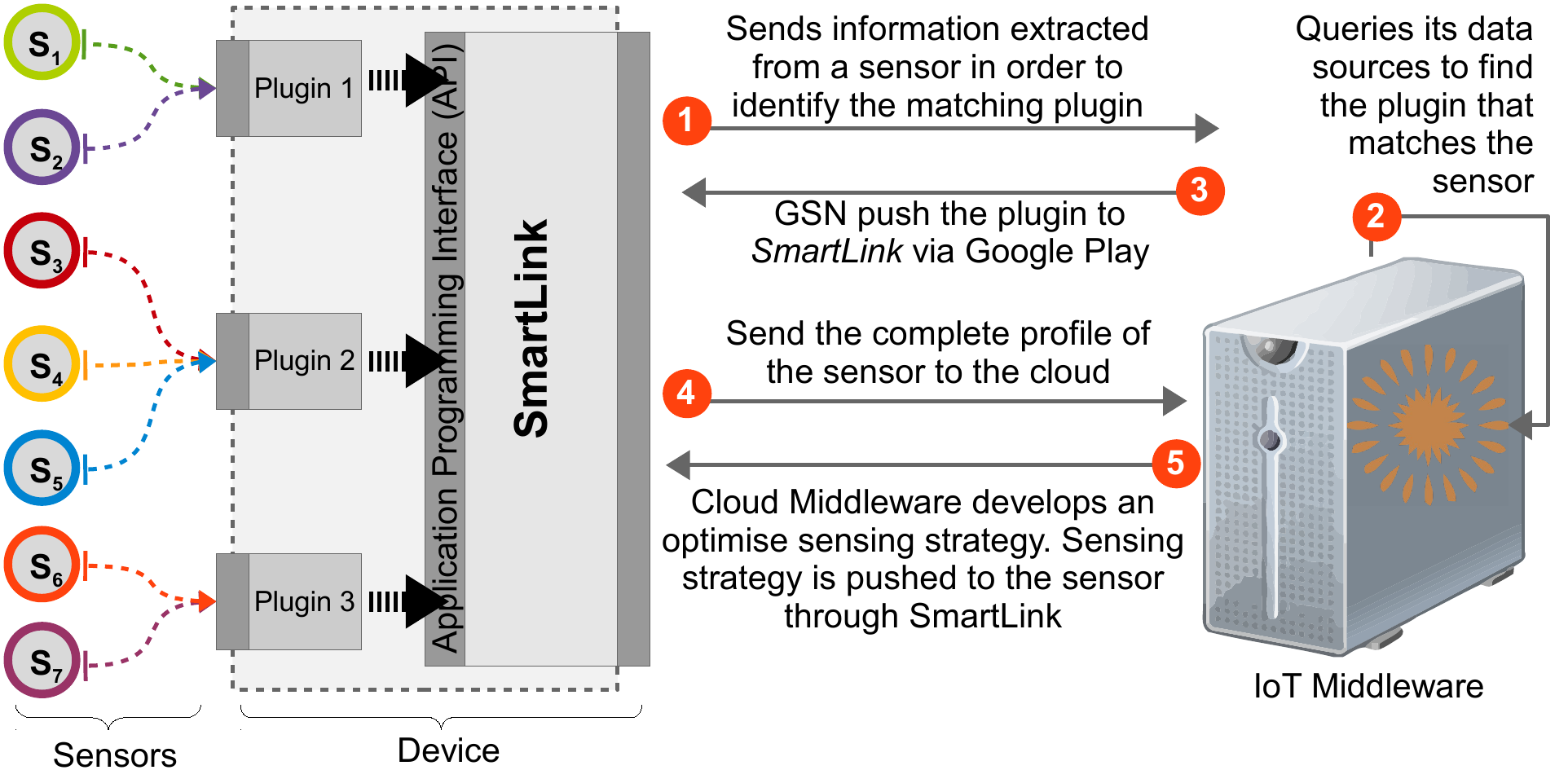}
 \caption{System architecture of the CADDOT model which consists of three main components: sensors, \textit{SmartLink} tool, and the cloud middleware. Order of the interactions are numbered.}
 \label{Figure:System}	
\vspace{-0.43cm}	
\end{figure}

\textbf{System Architecture:} The CADDOT model consists of three main components: sensors, a mobile device (i.e. \textit{SmartLink}), and the cloud middleware. All three components need to work collectively in order to perform sensor discovery and configuration successfully. Figure  \ref{Figure:System} illustrates the interactions between the three components. The phases we explained earlier relating to the CADDOT model in Figure \ref{Figure:Model} can be seen in Figure \ref{Figure:System} as well. As we mentioned before, \textit{SmartLink} is based on plugin architecture. The core \textit{SmartLink} application cannot directly communicate with a given sensor. A plugin needs to act as a mediator between the sensor and the \textit{SmartLink} core application, as illustrated in Figure \ref{Figure:System}. The task of the mediator is to translate the commands back and forth. This means that in order to configure a specific sensor, the \textit{SmartLink} core application needs to employ a plugin that is compatible with both the \textit{SmartLink} application itself and the given sensor. The sensors can be programmed in different ways. We do not restrict the developers to one single sensor-level program design. In order to allow \textit{SmartLink} to communicate with a sensor which runs different program designs, developers need to develop a plugin that performs the command translations. Plugin designing is guided by an AIDL interface which explains the mandatory functionalities that needs to be implemented.


\section{Implementation}
\label{sec:Implementation}


\noindent\textbf{Hardware Setup:} We employed the Global Sensor Network (GSN) \cite{P227}  middleware as the cloud IoT platform and hosted it on a laptop with Intel Core i5 CPU and 4GB RAM during the proof of concept validation. However, our  CADDOT model can accommodate any other IoT middleware as well. We deployed the \textit{SmartLink} application in a Google Nexus 4 mobile phone (Qualcomm Snapdrago S4 Pro CPU and 2 GB RAM), which runs Android platform 4.2.2 (Jelly Bean). We deployed 52 sensors in the third floor of the CSIT building (\#108) at the Australian National University. All  sensors we employed in our experiments are manufactured by Libelium (libelium.com). The sensors senses wide variety of environment phenomenon, such as temperature, motion, stretch, humidity, presence and so on.   \textit{SmartLink} supports sensor discovery and configuration using both WiFi and Bluetooth. Other communication technologies such as ZigBee and RFID are supported through Libelium \textit{Expansion Radio Boards}.

\begin{figure}[t]
 \centering
    \vspace{-5pt}
 \includegraphics[scale=0.27]{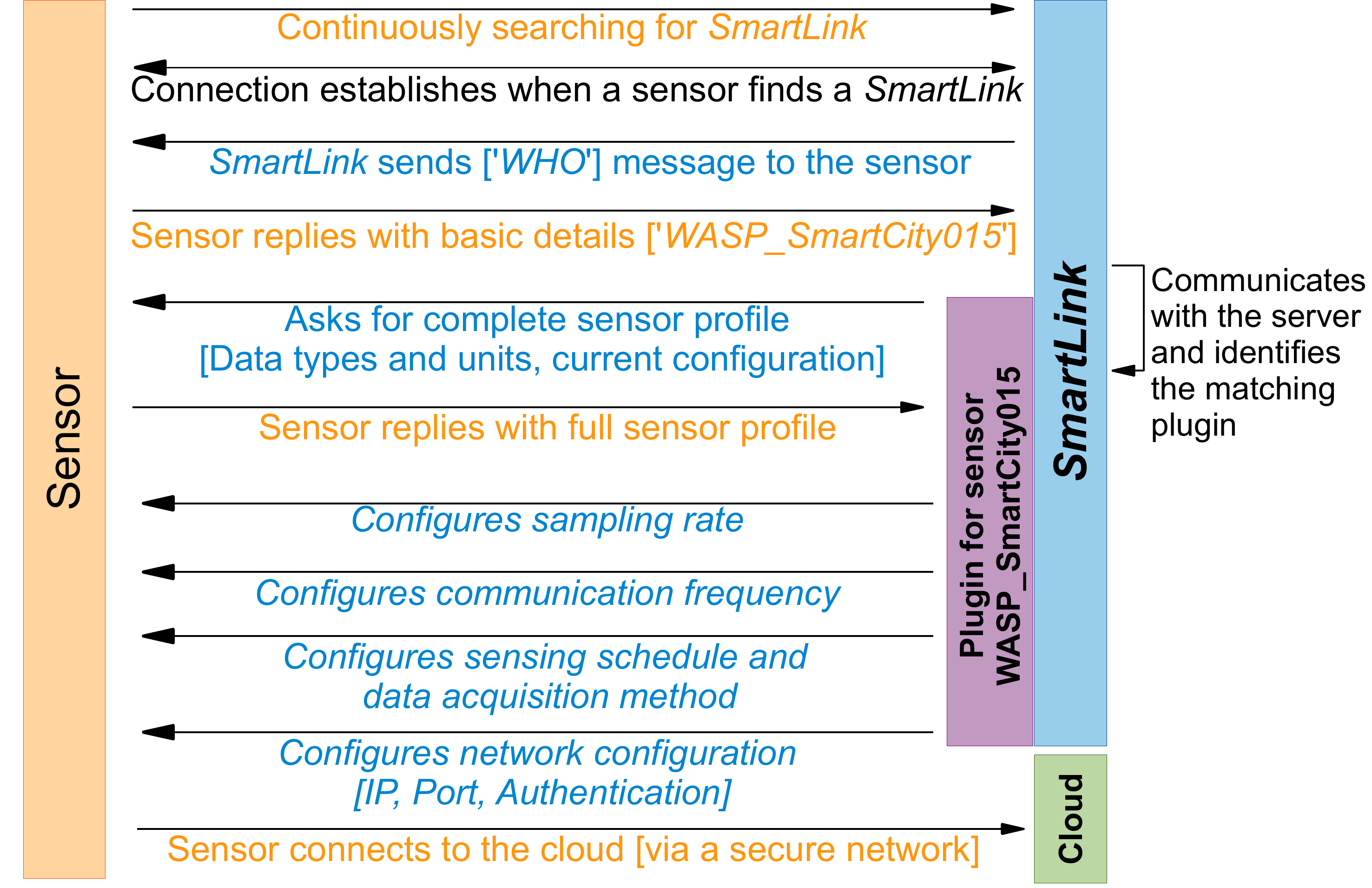}
  \vspace{-4pt}
 \caption{Sequence diagram demonstrates the interaction between a sensor and the \textit{SmartLink} tool. However, interaction sequence may varied depend on how the sensor is programmed.}
 \label{Figure:Sequence_Diagram}
 \vspace{-20pt}
\end{figure}

\noindent\textbf{Software Setup:} GSN \cite{P227} is developed in Java.  We extended GSN using the techniques proposed in \cite{ZMP005}, so it can configure it self accordingly. The Android platform has been used to develop the \textit{SmartLink} application.  Further, we employed the plug-in architecture called Android Interface Definition Language (AIDL) provided to facilitate plug and play functionality. AIDL allows \textit{SmartLink} to communicate with sensors effective and efficient manner. In order to simulate the heterogeneity of the sensors (in term of communication sequences), we programmed each sensor to behave and respond differently. As a result, each sensor can only be communicated by using a plug-in that supports the same communication sequence.  Figure \ref{Figure:Sequence_Diagram} shows how the interaction between sensor and the \textit{SmartLink} application occurs. We measure the average amount of time taken by each step (average of 30 sensor configurations). Figure \ref{Figure:Results1} illustrates the results and the following steps are considered: Time taken to (1) set up the sensor, (2) initiate connection between the sensor and \textit{SmartLink}, (3) initiate communication between sensor and \textit{SmartLink}, (4) extract sensor identification information, (5) retrieve the complete profile of the sensor, (6) configure the sampling rate, (7) configure the communication  frequency, (8) configure  the sensing schedule, (9) configure the network and authentication  details (so the sensor can directly connect to the cloud), (10) connect to the secure network using the provided authentication details.

\section{Evaluation}
\label{sec:Evaluation}

\begin{figure}[h]
 \centering
 \includegraphics[scale=0.4]{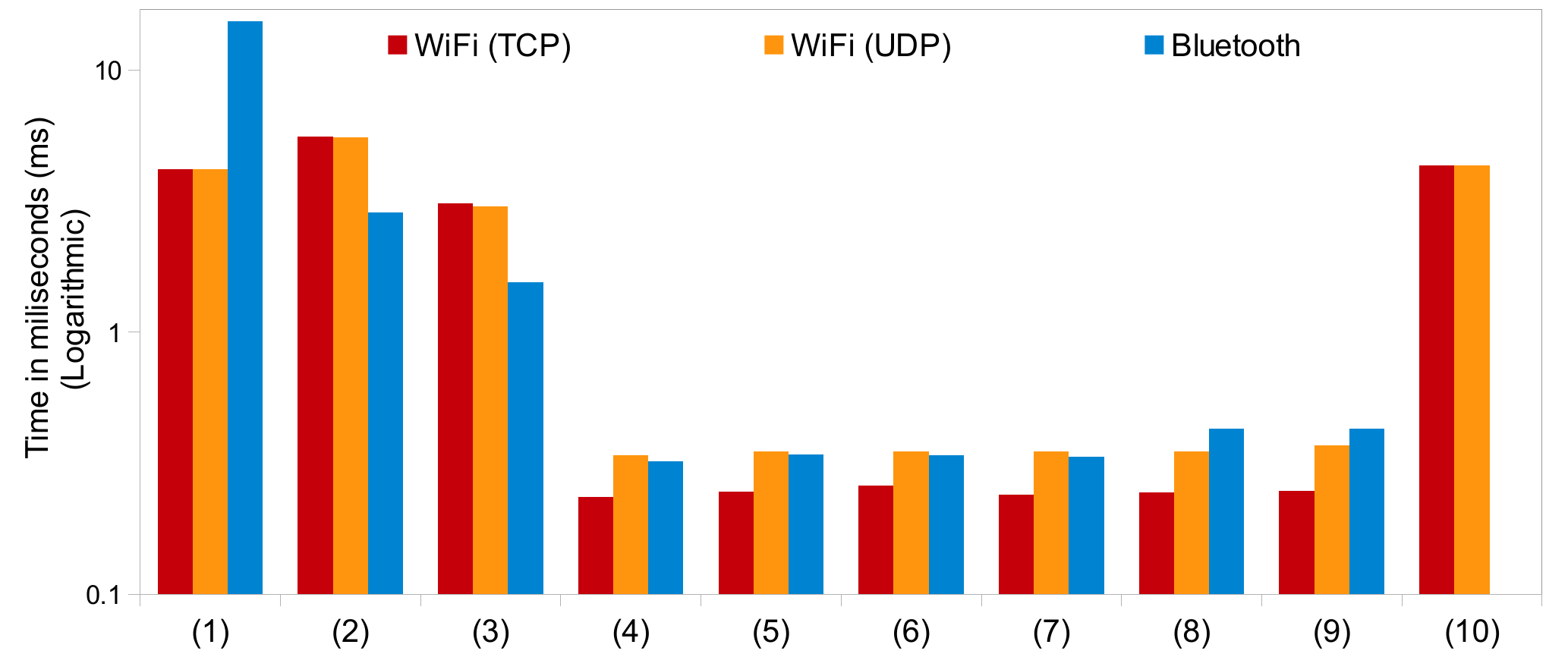}
  \vspace{-8pt}
 \caption{Time taken to configure a sensor in step-by-step. Experiments are conducted using TCP, UDP, and Bluetooth.}
 \label{Figure:Results1}
 \vspace{-10pt}
\end{figure}

According to Figure \ref{Figure:Results1}, the actual configuration tasks take less that one second. There is a slight variation in completion time in  configuration  steps (4) to (9). This is  due to storage access and differences in configuration command processing. Despite the protocol we use, sensors takes 5 to 15 seconds to boot and set-up themselves self. The setup stage consists of activities such as reading default configuration from storage, switch-on necessary modules and components (communication modules, real-time clock, SD card, sensor broads and so on).  According to the results, it is evident that a single sensor can be configured in less than 12 seconds (i.e. assuming sensors are already booted, which takes additional 5 to 15 seconds depending on the communication protocol).  Additionally, \textit{SmartLink} can configure multiple sensors at given time in parallel.  This is a significant improvement over a manual labour intensive sensor configuration approaches.

\section{Related Work}
\label{sec:Related_Work}

In this section, we review some of the state-of-the-art solutions developed by in academia, as well as commercial business entities. Our review covers both mature and immature solutions proposed by start-up initiatives as well as large-scale projects. The proposed CADDOT model as well as the \textit{SmartLink} tool help to overcome some of the weaknesses in  existing solutions.

There are commercial solutions available in the market that have been developed by start-up IoT companies \cite{P596} and the research divisions of leading corporations. These solutions are either still under development or have completed only limited deployments in specialized environments (e.g. demos). We discuss some of the selected solutions based on their popularity. \textit{Ninja Blocks} (ninjablocks.com), \textit{Smart-Things} (smartthings.com), and \textit{Twine} (supermechanical.com) are commercial products that aim at building smart environments \cite{P596}. They use their own standards and protocols (open or closed) to communicate between their own software systems and sensor hardware components. The hardware sensors they use in their solutions can only be discovered by their own software systems. In contrast, our pluggable architecture can accommodate virtually any sensor. Further, our proposed model  facilitates different domains (e.g. indoor, outdoor) using different communication protocols and sequences. In addition, the CADDOT model can facilitate very high dynamicity and mobility. \textit{HomeOS}  \cite{P597} is a home automation operating system that simplifies the process of connecting devices together. Similar to our plugin architecture, \textit{HomeOS} is based on applications and drivers which are expected to be distributed via an on-line store called \textit{HomeStore} in the future. However, \textit{HomeOS} does not perform additional configuration tasks (e.g. scheduling, sampling rate, communication frequency) depending on the user requirements and context information. Further, our objective is to develop a model that can accommodate a wider range of domains by providing multiple alternative mechanisms, as discussed in section \ref{sec:Design_Decisions}. Hu et al. \cite{P339} have proposed a sensor configuration mechanism that uses the information stored in TEDS \cite{P258} and SensorML \cite{P256} specifications. Due to the unavailability and unpopularity of TEDS among sensor manufacturers, we use  TEDS-like mechanism, by establishing a  standard communication message formats, to extract information from a give sensors in \textit{phase 2} of the  CADDOT model.

Actinium \cite{P636} is a RESTful runtime container that provides Web-like scripting for low-end devices through a cloud. It encapsulates a given sensor device using a container that handles the communication between the sensor device and the software system by offering a set of standard interfaces for sensor configuration and life-cycle management. The Constrained Application Protocol (CoAP) has been used for communication. Pereira et al. \cite{P641} have also used CoAP and it provides a request/response interaction model between application end-points. It also supports built-in discovery of services and resources. However, for discovery to work, both the client (e.g. a sensor) and the server (e.g. the IoT platform) should support CoAP. However, most of the sensor manufacturers do not provide native support for such protocols. \textit{Dynamix} \cite{P627} is a plug-and-play context framework for Android, which automatically discovers, downloads, and installs the plugins needed for a given context sensing task. \textit{Dynamix} is a stand-alone application that tries to understand new environments using pluggable context discovery and reasoning mechanisms. Context discovery is the main functionality in \textit{Dynamix}. In contrast, our solution is focused on dynamic discovery and configuration of sensors in order to support a sensing as a service model in the IoT domain. We employ a pluggable architecture which is similar to the approach used in \textit{Dynamix}, in order to increase the scalability  and rapid extension development by third party developers. The Electronic Product Code (EPC)  \cite{P110} is designed as a universal identifier that provides a unique identity for every physical object anywhere in the world. EPC is supported by the CADDOT model as one way of identifying a given sensor. Sensor integration using IPv6 in building automation systems is discussed in \cite{P656}. Cubo et al. [12] have used a Devices Profile for Web Services\footnote{http://docs.oasis-open.org/ws-dd/ns/dpws/2009/01} (DPWS) to encapsulate both devices and services. DPWS defines a minimal set of implementation constraints to enable secure web service messaging, discovery, description, and eventing on resource-constrained devices. However, discovery is only possible if both ends (client and server) are DPWS-enabled.

\section{Conclusion and Future Work}
\label{sec:Conclusion}

We explored the barriers in deploying IoT solutions in order to build smart environments. We understood that sensor configuration is one of the major challenges. To address this, we presented the CADDOT model, an approach that automates the sensor discovery and configuration process in smart environments efficiently by considering key factors such as growing number of sensors, heterogeneity, on-demand schedules, and sampling rates, data acquisition methods, and dynamicity.  CADDOT also encourages non-technical users to adopt IoT solutions with ease towards building their own smart environments. In this work, we evaluated sensor configuration using three popular communication protocols. We validate the CADDOT model by deploying it in an office environment.  As CADDOT required minimum user involvement and technical expertise, it significantly reduces the time and cost involved in sensor discovery and configuration. In the future, we will explore the possibilities of developing an efficient technique to identify a given sensor using context information and probabilistic techniques  in circumstances where information extracted in step 2 in CADDOT model is not adequate.



\textbf{Acknowledgements:} The authors acknowledge support from the OpenIoT Project, FP7-ICT-2011-7-287305-OpenIoT.





%
  \bibliography{Bibliography}
  \bibliographystyle{abbrv}

\end{document}